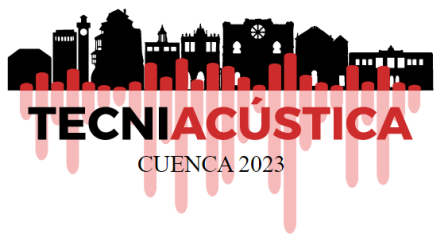

# DEL VISUAL AL AUDITIVO: SONORIZACIÓN DE ESCENAS GUIADA POR IMAGEN

*M. Sánchez-Ruiz[1\*], L. Fernández-Galindo[1], J.D. Arias-Londoño[1], M. Cámara-Largo[1,2], G. Comini[3],*
*A. Gabrys[3], J.L. Blanco-Murillo[1,2], J.I. Godino-Llorente[1], L.A. Hernández-Gómez[1,2]*

[1] ETSI Telecomunicación, Universidad Politécnica de Madrid. Avda. Complutense 30, 28040 Madrid
[2] Information Processing and Telecommunication Center, UPM. Avda. Complutense 30, 28040 Madrid
[3] Amazon.com

**RESUMEN**

Los recientes avances en las técnicas generativas de imagen, vídeo, texto y audio, y su uso por el gran público, están propiciando nuevas formas de generación de contenidos. Habitualmente, se abordaba cada modalidad por separado, lo que plantea limitaciones. La sonorización automática de secuencias visuales es uno de los mayores desafíos de cara a la generación automática de contenidos multimodales.

Presentamos un flujo de procesamiento que, a partir de imágenes extraídas de vídeos, es capaz de sonorizarlas. Trabajamos con modelos pre-entrenados que emplean complejos codificadores, aprendizaje contrastivo, y múltiples modalidades, permitiendo representaciones complejas de las secuencias para su sonorización.

El esquema propuesto plantea distintas posibilidades para la asignación de los audios y el guiado por texto. Evaluamos el esquema sobre un *dataset* de marcos extraídos de un videojuego comercial y sonidos extraídos de la plataforma Freesound. Las pruebas subjetivas han evidenciado que el esquema propuesto es capaz de generar y asignar audios automática y convenientemente a las imágenes. Además, se adapta bien a las preferencias de los usuarios, y las métricas objetivas propuestas presentan una correlación elevada con las valoraciones subjetivas.

**ABSTRACT**

Recent advancements in generative techniques for images, videos, text, and audio have been widely embraced by the public, leading to innovative content-generation methods. Typically, these methods address each type of content separately, which has limitations for automatization. One of the game-changing challenges is to automatically add sound to video and image sequences.

We present a processing pipeline that can automatically sonorize video sequences using multimodal representations. We utilize pre-trained models with complex encoders and contrastive learning during training. These models combine various modalities, allowing for sophisticated representations of image sequences and sound.

Our framework offers multiple options for audio allocation and text-guided sonorization. We evaluate the framework using a dataset of frames from a commercial video game and sounds from the Freesound platform. Initial subjective tests demonstrate that the framework can effectively generate and assign audio to images, matching user preferences. The proposed objective metrics based on the multimodal representations show a strong correlation with user ratings and are entirely based on embedding metrics.

***Palabras Clave***— Inteligencia artificial, sonorización automática de escenas, representación multimodal, aprendizaje contrastivo, aprendizaje máquina.

## 1. INTRODUCCIÓN

Las últimas décadas han presenciado un avance sin igual en el desarrollo de la inteligencia artificial (IA). Recientemente una serie de hitos han alterado el desarrollo de estas tecnologías, así como su acceso para el gran público y el empleo que hacen de ellas. Frente a los reconocidos avances en imagen y texto, la generación realista de audio y vídeo todavía plantea importantes retos. Esto se debe a la complejidad de generar audio de alta calidad, que resulte natural al oído humano, mucho más sensible que el ojo. Además, los audios deben guardar relación con las imágenes [1], evocando o ajustándose a las imágenes según el caso, o complementando su contenido con información adicional, del contexto o del ambiente, por ejemplo.

Mientras que existen multitud de plataformas capaces de generar imágenes a partir de descripciones, tales como DALL-E de OpenAI [2], hay muchas menos capaces de generar audio de calidad a partir de imágenes o descripciones [3, 4]. Los motivos detrás de esta situación podrían estar relacionados con la falta de grandes conjuntos de datos que incluyan pares de texto y audio de alta calidad [5], la falta de métodos que permitan un manejo eficiente [6], así como la complejidad de modelar los distintos niveles semánticos en el audio, desde el sonido ambiente, hasta los eventos sonoros o los efectos de audio. En este trabajo profundizamos en el reto de representar escenas para facilitar la identificación, o en su caso la generación, de registros de audio acordes a estas escenas y a sus representaciones.

Las nuevas funcionalidades de representación, tratamiento y generación de audio pueden permitirnos combinar varias modalidades y pensar en representaciones integrales, pasando de unas a otras de forma natural. Sin embargo, esto requiere superar el conocido "hueco de la modalidad" (*modality gap*), por lo que el procedimiento no resulta inmediato ni obvio. Además, los continuos avances en IA están favoreciendo la integración de distintas modalidades y sus combinaciones, así como la continua publicación de nuevos modelos entrenados. El trabajo con estas combinaciones de modalidades puede ser complejo y no siempre resulta sencillo monitorizar y asegurar un resultado de calidad.

La sonorización automática de escenas y secuencias de vídeo tiene multitud de aplicaciones dentro de la industria de la generación de contenidos [7], los videojuegos [8], la pre y la posproducción [9, 10], el desarrollo de ambientes inmersivos o en el cine [11], entre otros.

Este trabajo presenta un esquema automático de sonorización de escenas guiado por imagen que emplea algunos de los avances más recientes en inteligencia artificial. El esquema propuesto permite emplear tanto representaciones textuales como elemento intermedio (imagen a texto, y texto a audio), apoyándose en los modernos esquemas multimodales, como representaciones multimodales directas (imagen a audio).

Además de abordar la sonorización, el trabajo evalúa los resultados obtenidos en dos aspectos fundamentales. Por una parte, los sonidos deben ser de alta calidad, evitando efectos espurios y niveles de distorsión o de ruido que puedan resultar incómodos. Por otra parte, debe existir correspondencia y coherencia entre las distintas modalidades. En este sentido, hoy existen muy pocas métricas destinadas a este segundo análisis, y los autores no han identificado ningún trabajo anterior centrado en la sonorización de secuencias.

El resto del trabajo se ha estructurado de la siguiente manera. En la sección 2 se presenta el estado del arte, haciendo hincapié en las técnicas y modelos de representación multimodal, el aprendizaje contrastivo y las técnicas generativas. En la sección 3 presentamos el procedimiento de sonorización automática propuesto, desgranamos los detalles de cada uno de los módulos, del proceso de evaluación y de posibles alternativas sobre la estructura propuesta. En la sección 4 presentamos los materiales empleados para evaluar su funcionamiento, incluyendo los conjuntos de datos y los modelos empleados. En la sección 5 presentamos los experimentos diseñados, y en la 6 sus resultados. Las conclusiones y trabajos futuros se detallan en la sección 7.

## 2. ESTADO DEL ARTE

A lo largo de los últimos años los sistemas automáticos de aprendizaje máquina han alcanzado cotas muy avanzadas en el manejo de representaciones. Particularmente, los esquemas profundos logran captar distintos niveles de representación y significación, conectando las distintas modalidades de los datos (imagen, texto, audio), empleando formas análogas para todas ellas [16]. Los llamados y conocidos *embeddings* son representaciones matemáticas a modo de vectores de números con múltiples dimensiones (habitualmente cientos o miles) que encierran en ellos información útil sobre los datos empleando representaciones más compactas (menor tamaño que los datos originales) y operables. Los llamados codificadores son componentes centrados en una o varias modalidades y entrenados para calcular estos *embeddings*. Su empleo nos permite trabajar en espacios de números con propiedades matemáticas, en lugar de hacerlo sobre píxeles de imagen o los valores de las muestras de audio.

En contrapartida, también se entrenan decodificadores, que nos devuelven del espacio numérico de los *embeddings* a las representaciones modales. Los esquemas convencionales obligan a que codificación y decodificación se desarrollen entre espacios de la misma naturaleza (imágenes, audio o texto). Sin embargo, los multimodales permiten transitar entre modalidades, pasando de imagen a texto, y de texto a audio, o directamente de la imagen al audio, por ejemplo.

La Fig. 1 muestra un esquema de codificación con dos modalidades, imagen (izquierda) y audio (derecha), en este caso. Cada una emplea un esquema de codificación y ofrece un vector de salida para sus $N$ datos de entrada. En el centro de la figura superior aparece el aprendizaje contrastivo (*contrastive learning*), como estrategia para generar representaciones coherentes a las distintas modalidades. Este proceso de aprendizaje emplea datos no etiquetados y se centra en contrastar casos favorables frente a casos opuestos, permitiendo que el modelo aprenda diferencias intrínsecas entre las clases subyacentes.

Sin embargo, una de las claves del aprendizaje contrastivo es que de forma natural se observan diferencias intrínsecas en las representaciones de las distintas modalidades.

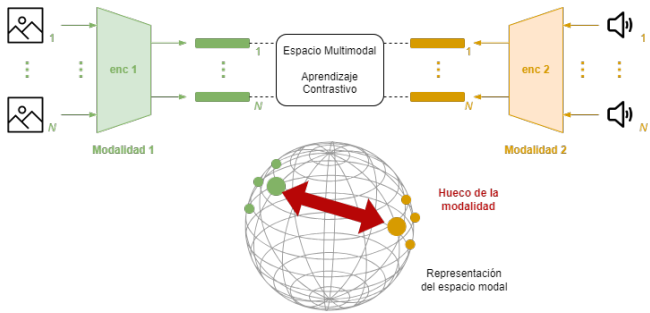

**Figura 1** Esquema general de la codificación de dos modalidades. El resultado del aprendizaje contrastivo es una representación coherente entre ambas con un hueco entre los subespacios de las modalidades.

Este es el llamado *hueco de la modalidad* [12], que da nombre al vacío que se origina entre los subespacios de representación de las distintas modalidades. Así, las representaciones obtenidas para las distintas modalidades no coinciden perfectamente, ni su correspondencia es directa. Esto se ilustra en la parte inferior de la Fig. 1 donde se encuentran representados los vectores de datos dentro de un espacio de dimensión más elevada. La visualización emplea una esfera, emulando la hiperesfera que describe el espacio *N*-dimensional de los *embeddings* según reporta la literatura para algunos modelos empleados en este trabajo [13].

Hoy día existen multitud de esquemas de representación que emplean el aprendizaje contrastivo, o que integran módulos que hacen uso de él. Entre los ejemplos más representativos figuran CLIP (*Contrastive Language-Image Pre-training*) para imagen y texto [14], CLAP (*Contrastive Language-Audio Pre-training*), para texto y audio [15], o ImageBind, que integra representaciones para imagen, texto y audio [16]. Estos módulos integran esquemas de codificación coherentes entre sí, que evitan la necesidad de decodificar. Hacerlo así permite trabajar en el espacio de representación y operar en él, pero no nos ofrece una forma de recuperar los datos.

Son muchas las aplicaciones que se benefician de manejar estas versiones codificadas, y muy pocas las que nos permiten limitarnos a operar con ellos sin volver al espacio natural de los datos. En este trabajo, el proceso de sonorización guiado por imagen exige tomar imágenes a la entrada y entregar audio a la salida. Asimismo, se requiere evaluar los resultados a la vista de las entradas recibidas. Esto se ilustra en la Fig. 2.

Existen multitud de esquemas que generan representaciones modales en distintos espacios. En particular, CoCa (*Contrastive Captioners*) es un módulo diseñado para obtener descripciones textuales breves del contenido de imágenes [17]. Esto es, permite pasar de imagen a texto, empleando internamente las representaciones anteriores.

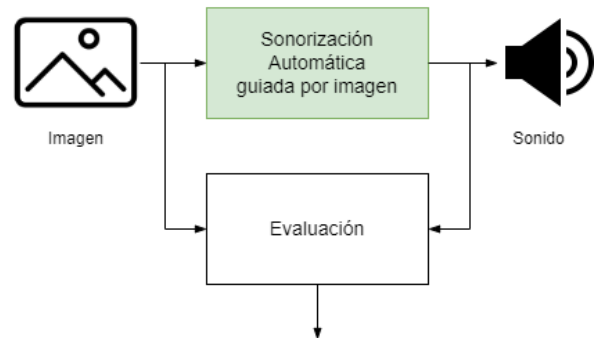

**Figura 2** Esquema de la sonorización automática guiada por imagen, y de la evaluación de resultados.

A este se suman otros módulos, como los capaces de generar audio a partir de representaciones textuales. Un ejemplo de esquema generativo capaz de hacer esto es AudioLDM (*Audio Generation with Latent Diffusion Models*), que emplea modelos de difusión [18]. AudioLDM destaca por su capacidad para generar audio a partir de representaciones textuales, aprovechando modelos de difusión, lo que garantiza una alta calidad en la síntesis de voz. Su uso de modelos de difusión latente (*Latent Diffusion Models*) proporciona una mayor flexibilidad y realismo en la generación de audio a partir de texto.

Una de las preguntas más relevantes entorno a estos modelos generativos es la evaluación de su calidad y su coherencia. Dado un dato representado en su modalidad correspondiente (imagen, texto o audio) no es sencillo evaluar su calidad. La Fig. 2 ilustra una cuestión central en la evaluación de estos modelos generativos, y es la necesidad de definir la calidad objetiva con relación al grado de concordancia entre la imagen y el sonido, lo que plantea desafíos en la medición precisa de esta coherencia. Esta discusión se entrelaza con la necesidad de desarrollar métodos generativos efectivos que logren mantener un alto nivel de fidelidad y correlación entre los datos de diferentes modalidades.

Más allá de todo esto, en este trabajo enfrentamos un problema más a raíz del hueco entre modalidades. Anticipamos que la coherencia entre modalidades en la evaluación y validación de resultados plantea un importante desafío, que tiene su origen en la brecha existente entre las distintas representaciones modales.

Con todo esto, ya tenemos todas las piezas que integrarán los esquemas de sonorización de este trabajo. La Fig. 2 ilustra el doble reto que enfrentamos. De una parte, diseñamos el proceso de sonorización (arriba), que puede emplear módulos como los referidos; y de otra, definimos una evaluación que permita analizar sus resultados, a la vista de las entradas y las salidas (abajo). En este trabajo presentamos un esquema que nos permite afrontar ambos retos a la vez.

### 3. EL PROCEDIMIENTO DE SONORIZACIÓN AUTOMÁTICA

En este trabajo hemos considerado dos esquemas diferentes. Estos se ilustran en la Figura 3. Ambos esquemas se ajustan perfectamente a las necesidades anteriormente descritas, desde dos aproximaciones marcadamente diferentes. El de la arriba (Esquema 1), emplea un esquema puramente guiado por imagen y asigna audios ya disponibles en una base de datos, tal como lo haría un sistema de recuperación (*retrieval*) automático. Estos audios se tratan previamente y representan convenientemente, de forma que el proceso de sonorización se limita a asignar el audio más conveniente a cada imagen. El proceso requiere de un esquema de representación capaz de combinar imagen y audio (en este caso se propone ImageBind), y de una métrica de ranquin para la asignación, que debe ser coherente con la métrica de entrenamiento de las representaciones. En este caso, emplearemos la distancia y la similitud coseno.

El esquema de debajo en la Figura 3 (Esquema 2) se apoya en el texto y genera audio a partir de las representaciones textuales, que median entre la imagen y el audio. En este caso, empleamos CoCa, analizando la configuración más conveniente para obtener representaciones textuales del contenido (*prompts*) que resulten favorables para la correcta generación de sonidos. En este caso mediante AudioLDM. CoCa permite generar varias representaciones textuales, adaptando sus parámetros. AudioLDM permite generar variaciones de sonidos, con lo que, de facto, llegamos a un funcionamiento análogo al del Esquema 1, contando con la representación textual intermedia y capacidades generativas.

La motivación detrás de estos dos esquemas es simple. En primer lugar, ImageBind fue publicado muy recientemente, en mayo de 2023 [16] y ya permite alinear las distintas modalidades. Anteriormente esto no era posible y, sin embargo, el planteamiento general de la sonorización era igual. Además, el Esquema 1 carece de capacidades generativas y se centra en sonidos preexistentes. Esto añade un nivel de control adicional sobre el proceso de sonorización, pudiendo validar la calidad de los sonidos de antemano. Sin embargo, más allá de la sonorización, la originalidad y la novedad son aspectos muy relevantes en la generación de contenidos, y este esquema no los aborda. Por este motivo, en el Esquema 2 exploramos las capacidades generativas como vía para enriquecer la experiencia multimodal de los usuarios. En este sentido, el papel modulador del texto en el segundo de los esquemas es crucial para controlar el proceso de generación. Trabajos futuros podrán considerar alternativas generativas que eviten pasar por la modalidad textual, tomando otras posibles entradas, por ejemplo, *embeddings* de imágenes. Una posible forma pasa por eliminar algunas etapas de los modelos existentes y/o entrenar transformaciones o transcodificadores.

Por último, volviendo sobre el proceso de evaluación de los resultados, ilustrado en la Figura 2, debemos analizar con cuidado el reto que se nos plantea. Partiendo de las entradas, en primer lugar, debemos asumir que éstas contienen información visual útil. Esto es, que muestran elementos o escenas más o menos habituales, que pueden estar incluidas en un proceso previo de entrenamiento más o menos agnóstico, como el que se emplea para los modelos anteriormente listados. Esto supone que, de facto, debemos buscar un mínimo nivel de calidad en las imágenes que representan escenas y al dominio, para asegurar que los resultados puedan ser coherentes. Algo similar sucede con los sonidos asignados o generados. La calidad de estos es uno de los factores más relevantes en la experiencia de los usuarios, y debemos asegurar un nivel suficiente, pero también de cara al correcto funcionamiento de los codificadores.

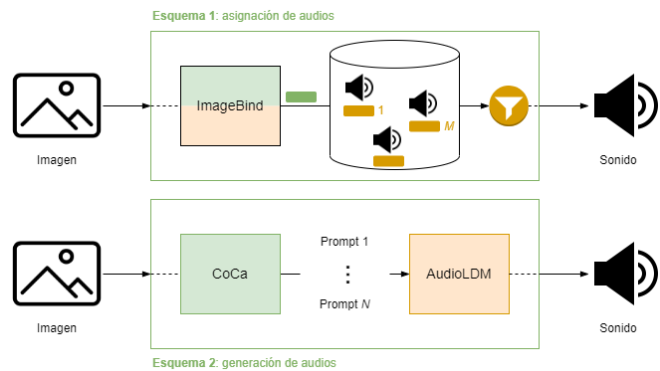

**Figura 3** Esquema de las implementaciones propuestas para sonorización automática guiada por imagen.

En cuanto a la coherencia o consistencia entre las entradas y las salidas, esto es, entre la imagen mostrada y el sonido asignado o generado, en la sonorización de escenas la opinión y valoración de los usuarios en un aspecto fundamental. Por ello, hemos recabado las valoraciones subjetivas de una muestra preliminar según la escala MOS (*Mean Opinion Score*). Éstas nos permitirán evaluar la correspondencia entre las sonorizaciones resultantes y las imágenes de partida, e identificar criterios objetivos capaces de aproximar las valoraciones de los usuarios. En este sentido este trabajo también plantea nuevas métricas, centradas en la consistencia entre los *embeddings* multimodales y en la métrica interna del espacio de representación, que comparamos con las valoraciones de los usuarios y la escala MOS.

La Fig. 4 resume las evaluaciones desarrolladas. La primera se centra en las evaluaciones subjetivas sobre la sonorización de distintas imágenes con distintos audios. Para la segunda y la tercera se emplea el espacio de representación de los *embeddings*, y se analiza tanto la consistencia en los *embeddings* como la distancia entre los mismos.

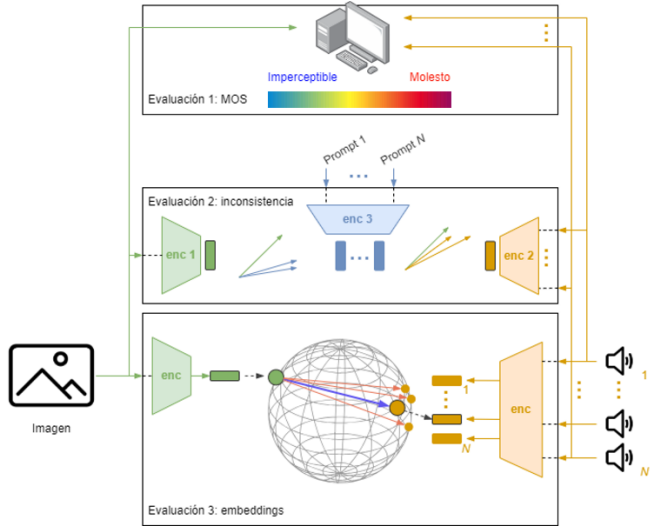

**Figura 4** Esquema sobre las Evaluaciones realizadas sobre el proceso de sonorización automática.

En concreto, la segunda evalúa diferencias en las direcciones, siendo capaz de tratar las tres distintas representaciones (o potencialmente dos, si fuera el caso). Para ello, no requiere de ninguna información más que los *embeddings*. La tercera, se centra específicamente en las entradas y las salidas del proceso de sonorización tal y como se define en la Fig. 1. Para ello, analiza la distancia entre la representación proyectada de la imagen sobre el subespacio de sonidos, tomando en consideración el hueco entre modalidades y trabajando sobre las proyecciones de las direcciones en los distintos subespacios de representación. Este cálculo requiere referencias en los respectivos subespacios, que permitan observar las diferencias locales y así tratar mejor las diferencias entre vectores.

Las ecuaciones (1) y (2) describen las expresiones empleadas en las evaluaciones objetivas, Evaluación 2 y Evaluación 3, respectivamente. La Ec. 1 emplea la distancia y la similitud coseno y considera tres vectores diferentes para analizar las distancias entre los *embeddings* de imagen y texto, $e_{imagen}$ y $e_{text}$, y los de imagen y audio, $e_{image}$ y $e_{audio}$.

$$\text{inc} = 2 \times \text{DisCos}(e_{image}, e_{text}) - \text{DisCos}(e_{image}, e_{audio}) \quad (1)$$

Como es bien conocido, la distancia coseno se deriva de la llamada similitud coseno, que toma valores entre -1 y 1: +1 cuando los vectores son idénticos, -1 cuando tienen sentidos opuestos y 0 si son vectores ortogonales. En consecuencia, la métrica de inconsistencia varía entre -1 y 2, donde 0 indica representaciones idénticas, 2 cuando imagen y texto son totalmente dispares, pero imagen y sonido coinciden, -1 cuando imagen y texto coinciden, pero imagen y audio son igualmente dispares, y +1 cuando las representaciones no coinciden entre sí.

Por otro lado, la ecuación Ec. 2 emplea la interpolación lineal esférica (SLERP, *spherical linear interpolation*) entre las representaciones para obtener una representación aproximada dentro del subespacio de otra modalidad.

$$e_{1audio\_t} = e_{0audio} + \text{SLERP}(e_{1marco}, e_{0marco}; \theta) \quad (2)$$

Aquí, $e_{i\text{-audio\_t}}$ se refiere al *embedding* del audio objetivo que correspondería al sonido deseable para sonorizar la imagen, $e_{0\text{-audio}}$ a un audio de referencia, $e_{j\text{-marco}}$ y $e_{0\text{-marco}}$ denotan las representaciones de las imágenes objetivas (las que se deben sonorizar) y la de referencia, respectivamente. Finalmente, $\theta$ es un parámetro utilizado en la función de Interpolación Lineal Esférica (SLERP, por sus siglas en inglés). Representa el peso asignado en la proyección esférica a la primera dirección con respecto a la segunda. En nuestro caso lo fijaremos a 0,5. En suma, esta interpolación permite ajustar la representación de audio a partir de la dirección de variación en un subespacio (en este caso, los *embeddings* de imagen), que se aplica en otro subespacio (aquí, el de los audios).

Finalmente, es importante destacar que los esquemas de sonorización propuestos constituyen un primer diseño centrado en el contenido estático de los marcos que componen el vídeo de una escena. Plantean importantes simplificaciones respecto a un caso completo al centrarse en los elementos estáticos de los marcos. La literatura describe ampliamente cómo las representaciones empleadas no son capaces de describir cambios o movimientos, que habitualmente están asociados con la acción y que, por tanto, cargados de significado. Sobre esta misma idea, tampoco son capaces de tratar variaciones temporales, como el inicio o el fin de esas mismas acciones, lo que impide la detección o la sincronización de eventos. Recientemente se han publicado algunos trabajos que abordan la problemática de la representación de vídeo [19]. Estos podrán servir de base a futuras investigaciones sobre la sonorización de acciones.

## 4. MATERIALES

Para la evaluación del flujo de sonorización diseñado hemos empleado secuencias de vídeos extraídas de un videojuego y de un *dataset* público [20], y otros descargados de la plataforma Freesound [21], así como imágenes de pruebas extraídas de bases de datos disponibles en abierto [22].

Para estas pruebas elegimos el videojuego *The Elder Scrolls V: Skyrim*. Se trata de un ARPG (*Action Role-Playing Games*) del tipo mundo abierto desarrollado por Bethesda Game Studios, publicado por Bethesda Softworks en 2011 y que fue remasterizado en 2016, con nuevos gráficos y efectos renovados en versión multiplataforma. La historia se presenta en un universo con tintes, estética y narrativa medievales. Los efectos de sonido y la banda sonora son una referencia por su capacidad para complementar el contenido visual y para enriquecer la experiencia interactiva del juego.

El videojuego Skyrim ha sido estudiado en trabajos anteriores. Entre otros, Freitas [23] explora la relación de esta comunidad con la música, el sonido, la inmersión y la modificación de juegos en la serie de Elder Scrolls. Sobre esta base, investiga cómo las percepciones de los usuarios sobre la música en Skyrim y otros títulos de Elder Scrolls pueden haber influido en su compromiso personal con los juegos y con experiencias cada vez más inmersivas. Estas ideas refuerzan nuestro interés en la sonorización automática.

El *dataset* que se emplea en [20] contiene vídeos de 5 segundos de 10 categorías diferentes: tiro con arco, braza (estilo de natación), ballesta, baile, esquivar, vuelo, montar a caballo, paracaidismo, y agitar el arma. Hasta cinco de éstas, tiro con arco, montar a caballo, correr, braza y paracaidismo, se incluyen en el conjunto de datos UCF101 [24]. Desafortunadamente, cada categoría cuenta solo con 10 vídeos de 5 segundos cada uno y con una resolución de 320x180 a 30 cuadros por segundo. Por tanto, la calidad de imagen es baja y los materiales son muy limitados.

En vista de todo lo anterior, para este trabajo hemos optado por diseñar un nuevo *dataset* siguiendo los mismos criterios del trabajo anterior. En este caso, se han buscado vídeos de alta resolución y se han definido escenas que difieren no sólo en las acciones del personaje y publicados en plataformas en abierto. La Fig. 5 muestra ejemplos de las imágenes empleadas en las pruebas. Las imágenes a) y b) representan ejemplos de escenas diferentes del *dataset* de baja resolución original, mientras que c) y d) corresponden a escenas del nuevo. Por último, e) y f) son ejemplos tomados de fuera del dominio, y que contrastan con las anteriores por su contenido visual. Estas últimas las emplearemos al evaluar la capacidad del esquema de sonorización para distinguir muestras que se alejan de nuestro dominio de trabajo. En ese caso, esperamos que sus *embeddings* se sitúen en regiones alejadas de las correspondientes a las escenas a sonorizar. Eso mismo debe suceder con las representaciones de los sonidos, en función del grado de concordancia entre sus contenidos.

Sobre las secuencias a sonorizar, se extrajeron 18068 marcos de alta resolución (Full-HD, 1920x1080 píxeles) empleando la herramienta FFMPEG [25]. Se agruparon manualmente en once escenas diferentes atendiendo a su contenido. Además, se analizaron fragmentos de las escenas entorno a una imagen por referencia seleccionada para para la generación de textos alternativos. Se tomaron 10 marcos por delante y por detrás, trabajando sobre 21 imágenes, lo que finalmente arrojó un total de 4620 textos y se generaron sus correspondientes 4620 audios, uno por marco. En las pruebas preliminares realizadas sobre un único texto por imagen, este número de textos y registros de audio se redujo correspondientemente.

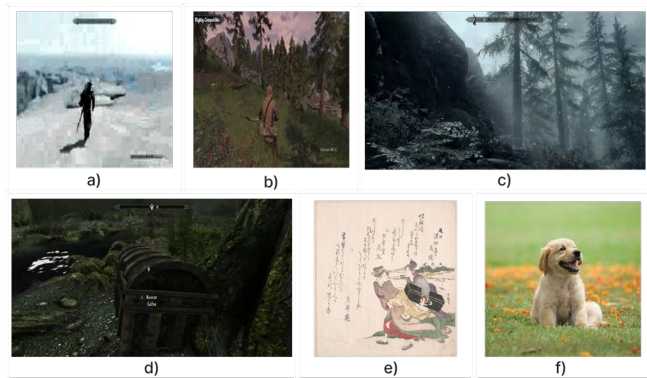

**Figura 5** Capturas extraídas del *dataset* de Skyrim: a) y b) de baja resolución en escenas distintas, mientras c) y d) pertenecen al de alta resolución, en las escenas "Caminando por el bosque" y "Abriendo un cofre", respectivamente. Por último, e) y f) son ejemplos tomados de otros dominios.

En cuanto a los audios para la evaluación, se extrajeron 100 ficheros de audio de la plataforma Freesound, realizando consultas semiautomáticas sobre términos preestablecidos, a la vista de la temática general del videojuego y del contenido visual de las escenas seleccionadas. Se trata de un conjunto limitado, pero que debe ser suficiente para evaluar el concepto de sonorización que presenta este trabajo.

## 5. EXPERIMENTOS

A continuación, describimos los cuatro experimentos desarrollados para la evaluación de la sonorización automática de las escenas de videojuego guiada por imagen. El proceso descrito sigue los pasos que ilustra la Fig. 4, empezando por la evaluación subjetiva sobre valoraciones de usuarios, la evaluación objetiva de la incoherencia a partir de la distancia coseno entre las distintas representaciones, y la evaluación objetiva a través de las proyecciones.

## 5.1. Evaluación subjetiva

Esta evaluación involucra participantes humanos que asignan calificaciones según la escala MOS (1 muy pobre, 5 excelente) para indicar su percepción subjetiva de la fidelidad del audio respecto a la imagen fuente. Se han seleccionado tres escenas del *dataset* de Skyrim: "Caminando por el bosque", "Paseo junto al río" y una escena de "Abriendo un cofre". De éstas, se extrajeron 63 cuadros y se generaron 63 sonidos, uno para cada cuadro, para la evaluación subjetiva.

La evaluación contó con siete evaluadores que recibieron recomendaciones básicas como el uso de auriculares, escuchar el audio de referencia para cada escena y familiarizarse con el sonido deseado, comparar el audio de referencia con cualquier audio generado para esa escena, y tomarse el tiempo suficiente para considerar cada muestra. Con todo, cada evaluación asignó una puntuación MOS por cada pareja cuadro-audio que le fue presentado.

## 5.2. Evaluación objetiva: distancia coseno

Esta prueba analiza las distancias entre *embeddings* de las diferentes modalidades involucradas. Buscamos cuantificar las disparidades entre las representaciones extremo a extremo (imagen a audio, según Esquemas 1 y 2 en Fig. 3) y analizar el papel mediador del texto (Esquema 2, misma figura). Proponemos la métrica de inconsistencia de la Ec. 1. De esta forma, podemos corroborar si las imágenes de entrada y los audios de salida ofrecen representaciones consistentes, que se asemejan entre sí, mientras evaluamos también la consistencia con el texto, para así identificar problemas subyacentes en las modalidades que afecten al esquema de sonorización propuesto. Además, esta forma de trabajar nos permite ordenar los resultados y seleccionar los ejemplos con menor distancia en absoluto para tener la mejor sonorización.

Empleamos las bases de datos de Skyrim, de alta y baja resolución y codificamos las representaciones con ImageBind, incluidas las correspondientes a 12000 subtítulos de CoCa y 12000 audios generados por AudioLDM: 20 descripciones y 20 audios asociados por cada marco. Calculando los valores de la métrica según Ec. 1, tomamos $k=10$ para evaluar los mejores resultados.

## 5.3. Evaluación objetiva: SLERP

Este enfoque permite tanto evaluar cómo distintas salidas de audio se asemejan a las referencias y así sonorizar una escena (Esquema 1 en Fig. 3), como una evaluación integral de la sonorización extremo a extremo y (también del Esquema 2 de la Figura). En este caso, empleamos ImageBind y la Ec. 2.

La función SLERP se ha evaluado tanto con fotogramas que pertenecen a la misma escena como con fotogramas de escenas diferentes. En ambos casos, se emplea un valor de $\theta$ igual a 0,5. Además, al evaluar la proximidad entre representaciones se evaluaron las distancias coseno y euclídea (Ec. 3). Esto permite estudiar las diferencias entre las representaciones tanto en magnitud como en dirección.

$$\text{Distancia SLERP} = \text{dist}(e_{1\text{-audio}}, e_{1\text{-audio\_t}}) \qquad (3)$$

El resultado de esta función es un número que representa la distancia entre *embeddings*. El valor mínimo que puede obtener es cero, lo que indica que los *embeddings* son idénticos, esto es, que se superponen. Además, conforme a la Ec. 3 podemos ranquear los *embeddings* de los audios, considerando el más apropiado aquel, de entre un conjunto de posibles, que muestre la menor distancia.

Para esta evaluación empleamos la base de datos de alta resolución de Skyrim. Disponemos de 10 audios de referencia, los 11 marcos de referencia (uno por escena), las 231 marcos seleccionadas (las 10 anteriores y los 10 siguientes, respecto al de referencia), así como de los 4000 audios generados.

## 5.4. Estudio de la relación entre los distintos resultados de evaluación

Finalmente, analizamos los resultados de las evaluaciones subjetivas de los usuarios frente a los resultados obtenidos con las métricas objetivas anteriores. El objetivo es cuantificar la correlación entre ambas tomando como referencia las evaluaciones de los usuarios, y empleando la métrica objetiva para aproximar sus valoraciones.

Para ello, disponemos de más de 400 valoraciones, realizadas por de siete evaluadores distintos. Cada evaluación se compone de una terna de imagen, sonido, y una puntuación, acotada entre 5 (excelente) y 1 (malo), con cinco niveles. A estas ternas, les incorporamos los resultados de cada una de las evaluaciones objetivas. Para nuestro análisis estadístico, tomamos cada una de estas valoraciones como independientes, sin asumir nada sobre su distribución, y nos centramos en el coeficiente de correlación de Pearson entre las métricas previas y las valoraciones. En esta ocasión, evitamos considerar ningún tipo de modelo de relación entre las evaluaciones. Tampoco se han segmentado las entradas, ni las valoraciones. Trabajos futuros podrán considerar todas estas estrategias, así como modelos más complejos y entrenados para predecir unas evaluaciones a partir de otras.

## 6. RESULTADOS

Las figuras siguientes presentan los resultados de diversas evaluaciones en el orden en que se llevaron a cabo, como se detalla en la sección 5. En particular, la Figura 6 muestra las evaluaciones subjetivas utilizando la escala MOS para tres escenas diferentes y los 21 audios evaluados en el Experimento 1.

Basándonos en la Figura 6, observamos lo siguiente:

1. Los evaluadores están generalmente de acuerdo en que el audio de la escena "Caminando por el bosque" es de alta calidad, con calificaciones por encima de 4.
2. Para la escena "Paseo junto al río", también hay un consenso entre los evaluadores sobre la buena calidad del audio.
3. Sin embargo, el audio de la escena "Abriendo un cofre" recibe calificaciones más bajas, por debajo del valor 3. Esto podría deberse a que el modelo AudioLDM no fue entrenado con sonidos que se puedan asociar con la acción de abrir un cofre, lo que resulta en sonidos generados que se desvían de las expectativas.

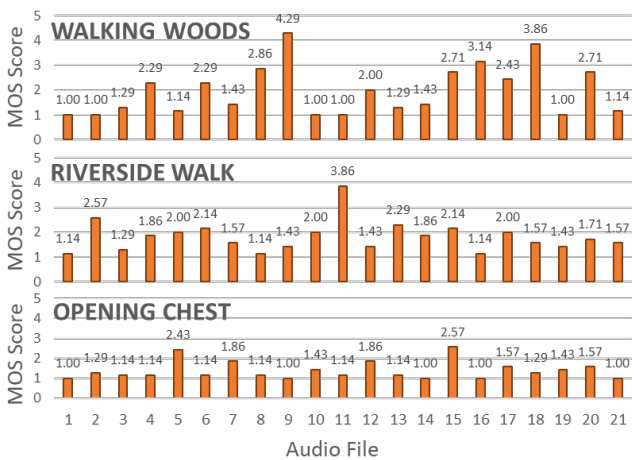

**Figura 6** Resultado de la Evaluación 1 para las tres escenas.

Otra forma de analizar los resultados permite considerar la distribución estadística de las valoraciones. En la tercera escena todos los valores son bajos, mientras que en las dos primeras un subconjunto de los registros obtuvo buenas valoraciones respecto al resto. Un estudio más amplio requerirá incrementar el tamaño del conjunto de audios a considerar, aumentando su riqueza y variabilidad. Asimismo, cabe considerar registros con distintos niveles de calidad, a fin de asegurar que la sonorización cubra las expectativas de los usuarios.

En cuanto a la Evaluación 2, al calcular la distancia del coseno para la métrica de Inconsistencia, Ec. 2, los resultados de las distancias entre los *embeddings* de imagen y texto (ver Fig. 7.a) y los *embeddings* de imagen y audio (ver Fig. 7.b) se acercan a 1 para el *dataset* de baja resolución. Esto indica un alto grado de disimilitud o una relación casi opuesta entre los vectores que se están comparando. En otras palabras, no ha captado ninguna relación entre el contenido visual y el texto o audio asociados. Esto resulta lógico a la vista de la calidad de las imágenes en este conjunto ver Fig. 5.a y 5.b.

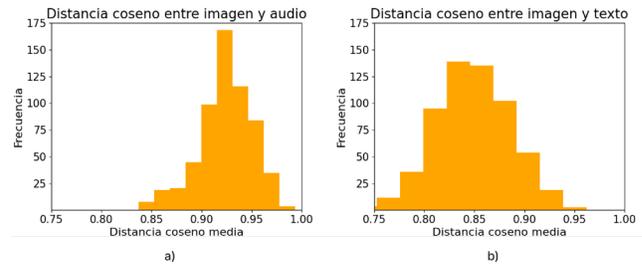

**Figura 7** Histogramas de distancias correspondientes a la Evaluación del *dataset* 2: baja resolución. Las diferencias son pequeñas y los valores bajos.

Vistas estas posibles limitaciones, se repitieron las pruebas sobre el conjunto de imágenes de alta resolución (Fig. 5c y 5d, por ejemplo), obteniéndose unos nuevos histogramas marcadamente diferentes a los de la Fig. 7. Estos se muestran en la Fig. 8.

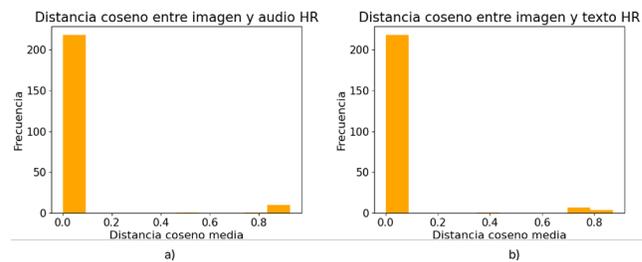

**Figura 8** Histogramas de distancias correspondientes a la Evaluación 2 para secuencias de alta resolución. Las diferencias son mayores y los valores más altos que en Fig. 7.

En lo que respecta a la tercera evaluación, se llevaron a cabo pruebas similares a las realizadas anteriormente. Sin embargo, en este caso, se centraron en las representaciones generadas para los archivos de audio y los fotogramas de las secuencias de imágenes, utilizando los codificadores modales de ImageBind.

**Tabla 1.** Tabla de la similitud entre pares modales correspondiente a la Evaluación 3.

| Par audio | Par imagen | Media | Desviación típica |
|---|---|---|---|
| No relacionado | Relacionado | 24,9 | 2,5 |
| No relacionado | No relacionado | 24,7 | 2,5 |
| Relacionado | No relacionado | 12,5 | 3,2 |
| Relacionado | Relacionado | 12,5 | 3,2 |

La tabla presenta una evaluación por modalidad (audio e imagen) de los pares referencia-objetivo, y evalúa la relación de similitud que existe entre ellos. Esta relación se basa en la

Ec. 2. En este contexto, los pares se dividen en dos categorías: "No relacionado" (objetivo alejado de la referencia) y "Relacionado" (objetivo próximo a la referencia).

Cuando los pares guardan relación, sus distancias se encuentran en el rango aproximado de 12,5. Esto sugiere similitud o relevancia entre los objetivos y sus referencias correspondientes. Por el contrario, los pares que no están relacionados con la referencia a la que se comparan presentan distancias mucho más elevadas, superando 24 unidades. Esto indica una notable disimilitud y falta de relevancia entre objetivos y referencias.

Por otro lado, cuando los pares son etiquetados como "Relacionado" (lo que implica que la imagen y el audio están próximos en sus representaciones multimodales), la métrica de Evaluación 3 arroja un valor medio de 12,5, con una desviación típica de 3,2. Esto sugiere que, en promedio, cuando existe una relación entre la imagen y el audio, la métrica produce un valor menor que en el caso de "No relacionado". Esta medición tiende a ser más variable, lo que podría reflejar la diversidad de grados de relación entre los elementos evaluados.

En última instancia, comparamos los resultados de las valoraciones subjetivas obtenidas en la Evaluación 1 con los datos objetivos de las Evaluaciones 2 y 3. Al igual que en las pruebas objetivas, los experimentos subjetivos abarcan todo el rango de resultados posibles. Esto respalda la idea de que la métrica objetiva y las percepciones subjetivas son coherentes. Finalmente, la Tabla 2 presenta el coeficiente de correlación de Pearson para cada uno de los experimentos subjetivos, así como su promedio.

**Tabla 2.** Coeficientes de correlación de Pearson de los experimentos subjetivos y la métrica objetiva de SLERP sobre los experimentos de la Evaluación 1.

| Exp. 1 | Exp. 2 | Exp. 3 | Media |
|---|---|---|---|
| -0,73 | -0,19 | -0,29 | -0,40 |

Basándonos en estos resultados, queda claro que hay una relación entre las métricas objetivas y las valoraciones subjetivas. En particular, se observa una correlación negativa entre la distancia de los *embeddings* y la evaluación subjetiva de los participantes. Planeamos expandir este estudio con más evaluaciones subjetivas para confirmar esta observación clave. Además, estos hallazgos respaldan la idea de dirigir futuras investigaciones hacia nuevas áreas, como estudios que comiencen con representaciones multimodales, en lugar de enfocarse únicamente en las representaciones modales iniciales.

## 7. CONCLUSIONES

Hemos logrado implementar y evaluar dos esquemas completos de sonorización automática empleando representaciones multimodales resultantes del aprendizaje contrastivo. Para ello, hemos integrado modelos pre-entrenados descritos en la literatura más reciente.

Además, hemos validado los resultados obtenidos mediante pruebas subjetivas desarrolladas con usuarios. Estos se han contrastado con las métricas objetivas propuestas, centradas en considerar la consistencia entre las imágenes ofrecidas y los audios generados o asignados, respectivamente. En conclusión, el esquema de sonorización propuesto es capaz de identificar los sonidos más apropiados para sonorizar una imagen de entre un conjunto dado, y generar audios coherentes con los marcos de video siempre que la resolución sea suficiente.

El código para calcular la métrica de distancia de incrustación, que mide la distancia entre un audio de referencia y un audio generado, está disponible en GitHub.

Como trabajos futuros, se está trabajando en el diseño de una página web que permita experimentar el esquema propuesta de forma libre, escuchar y evaluar los resultados, incorporando aspectos de la calidad del sonido (fidelidad, ruido, etc.). También se quiere evaluar la relación entre las métricas obtenidas y las métricas que emplea Freesound para ordenar la relación entre registros, y entrenar un esquema generativo completo que permita evitar la etapa intermedia de texto. Esto último va a requerir nuevas evaluaciones y experimentos, que permitan contrastar sus resultados conforme a las distintas métricas propuestas.

## 8. AGRADECIMIENTOS

Este trabajo ha sido financiado conjuntamente por el Ministerio de Economía y Competitividad del Gobierno de España dentro del proyecto PID2021-128469OB-I00, el Programa de Investigación e Innovación de la Unión Europea Horizon 2020 dentro del "Grant Agreement No. 101003750". Las becas de investigación de María Sánchez y Laura Fernández han sido financiadas por Amazon dentro de la colaboración IPTC-Amazon.

## 9. REFERENCIAS